\documentclass{PoS}
\title{{\small DESY 10--067 \newline SFB/CPP--10--40 \newline HEPTOOLS 10--018} \\[1.5cm]
Calculating multileg one-loop processes.
The case of $gg\rightarrow t \bar{t}gg$}

\ShortTitle{Calculating multileg one-loop processes}

\author{\speaker{Theodoros Diakonidis}, Bas Tausk \\
Deutsches Elektronen-Synchrotron
DESY, Platanenallee 6, 15738 Zeuthen, Germany\\
E-mail: \email{theodoros.diakonidis@desy.de}}
\abstract{This study is targeted to the NLO corrections of multileg
processes, very important for the LHC.
Starting from the construction of Feynman diagrams, the analytical
reduction of general one-loop integrals to scalar master ones, the
calculation of color structures, manipulation of spinor lines and
other amplitude constituents and finally phase space point selection
are obtained by use of a program producing Fortran code for numerical
calculation of one-loop corrections for processes like
$gg\rightarrow t \bar{t}gg$.}

\FullConference{13th International Workshop on Advanced Computing and Analysis
Techniques in Physics Research, ACAT2010\\
February 22-27, 2010\\
Jaipur, India}

\begin{document}

\section{\label{Intro} Introduction}
The era of LHC has finally arrived, after many years of waiting and
anticipating, and all the particle physics community is waiting
for the first results. There has been a lot of activity the last two years
in the branch of NLO calculation. With almost all interesting cases
of $2\rightarrow 3$ processes having been calculated, the interest has
moved towards $2\rightarrow 4$.
Processes like $pp\rightarrow t\bar{t}+b\bar{b}$
\cite{Bredenstein:2009aj,Bredenstein:2010rs,Bevilacqua:2009zn},
$V+3jet$ production
\cite{Berger:2009zg,KeithEllis:2009bu,Berger:2009ep,Berger:2010vm},
$qq\rightarrow b\bar{b}b\bar{b}$
\cite{Binoth:2009rv} and most recently the case of $pp\rightarrow
t\bar{t}+2 jets$ \cite{Bevilacqua:2010ve} have been studied in the last three
years, providing the background for future discoveries. This study is
also towards this direction.
\section {\label{rank1234}The method}
\begin{figure}[htbp]
\begin{center}
  \includegraphics[angle=90,scale=0.5]{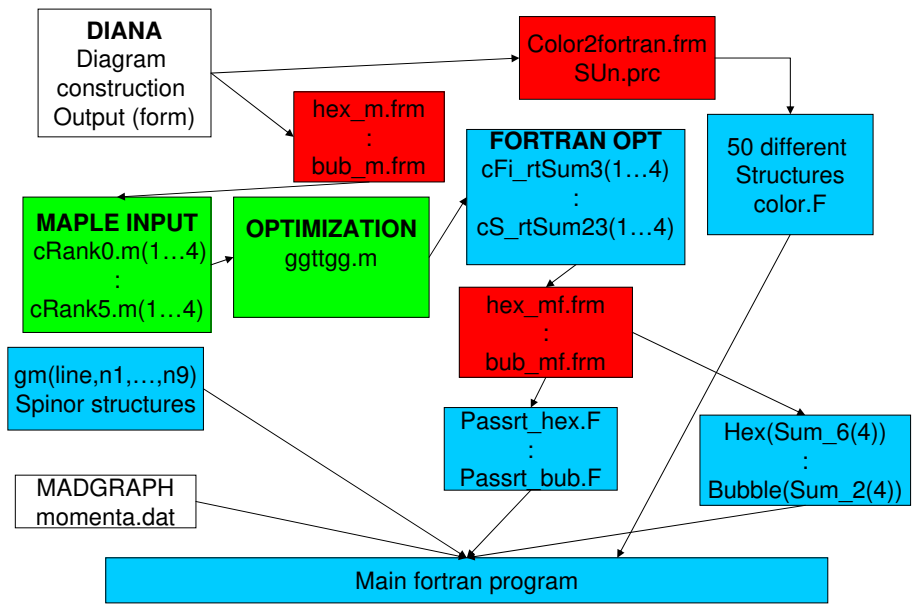}
  \caption{Graphical presentation of the case of $gg\rightarrow t \bar{t}gg$ }
  \label{fig:ggttgg}
\end{center}
\end{figure}

Our method is based on the Feynman diagrammatic approach. Which means
that we calculate the total amplitude by taking into consideration the
contributions of each of the one-loop Feynman diagrams that exist for
the NLO calculation. The final program calculating the amplitude is in
Fortran, although a variety of tools are used in combination (see Figure
\ref{fig:ggttgg}): Diana \cite{Tentyukov:1999is} for constructing the
diagrams, Form \cite{Vermaseren:2000nd}
for manipulation of the Diana output (the steps of calculation in Form
are colored red), Maple for optimization
of the produced Fortran code (green color) and finally Fortran for the
numerical evaluation (blue color). In the next session we will explain
in more detail the complete calculation, which ends by providing the
amplitude from an initial phase space point.

\subsection {The steps of the calculation}
As mentioned above, the first step is implemented by using Diana. We
provide the program with the appropriate Feynman rules and ask for the
construction of diagrams. For the case of $gg\rightarrow t{\bar{t}}gg$
the total number of diagrams is 4510.
Diana provides us with an output in Form which we further manipulate. The
next step is to strip the one-loop amplitude from the color structures.
In a few words using Form to manipulate the output of Diana we strip
off and manipulate the color structures using color Algebra. We
translate the Form result to Fortran code and end up with a Fortran
array containing all the possible color information.
For the case of $gg\rightarrow t{\bar{t}}gg$ there are 50 color structures,
which can be further divided in 4 main color groups. Representatives of
each group are shown in Table \ref{tab:ggttgg}. All other color structures
are produced from these ones through permutations of the external gluons.

\begin{table}[htb]
\centering
\begin{tabular}{|c|c|}
\hline
Color Structures & Number of cases \\
\hline
$(T^{\alpha_1}T^{\alpha_2}T^{\alpha_3}T^{\alpha_4})_{f_1f_2}$& 24\\
\hline
$Tr\{T^{\alpha_1}T^{\alpha_2}\}(T^{\alpha_3}T^{\alpha_4})_{f_1f_2}$& 12\\
\hline
$Tr\{T^{\alpha_1}T^{\alpha_2}T^{\alpha_3}\}(T^{\alpha_4})_{f_1f_2}$&8\\
\hline
$Tr\{T^{\alpha_1}T^{\alpha_2}T^{\alpha_3}T^{\alpha_4}\}\delta_{f_1f_2}$&6 \\
\hline
\end{tabular}
\caption[]{Color groups for the $gg\rightarrow t{\bar{t}}gg$ case.
The number of cases refers to all the possible permutations
of the gluons}\label{tab:ggttgg}
\end{table}

The next step is the manipulation of the color stripped amplitude. What
is left after the color bracketing is the spinor line connecting the
two external fermions $t {\bar{t}}$ with all the possible ways and the
rest are the tensor integrals and scalar products of external momenta
produced by propagators outside the loop and finally the polarization
vectors of the gluons.

The spinor line contribution is calculated by Fortran. After manipulating
it using all the possible contracting equations and simplifying the
result by moving the appropriate momenta to the corners of the spinors
and applying the Dirac equation, we end up with the simplest structure. Then
we glue the two spinors left and right and calculate the
result numerically in Fortran.

The tensor integrals are identified and flagged by Form. There is a
Fortran program that calculates them according to the New Recursive
Reduction Method of \cite{Diakonidis:2009fx}. The program called OLOTIC
(One LOop Tensor Integral Calculator), applies the formulas of the paper
expressing the tensor integrals in terms of scalar master integrals.
\begin{figure}[htbp]
\begin{center}
        \includegraphics[scale=0.5]{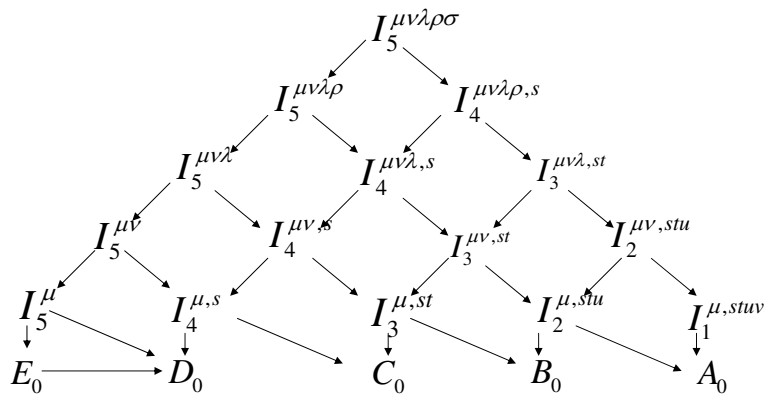}
        \caption{Graphical presentation of the tensor reduction procedure}
   \label{fig:reduction}
\end{center}
\end{figure}
In a few words, the idea of the reduction is that, one-loop $n$-point
rank $R$ tensor Feynman integrals [in short: $(n,R)$-integrals] for
$n\leq 6$ with $R\leq n$ are represented in terms of $(n,R-1)$-
and $(n-1,R-1)$-integrals (see e.g. Figure \ref{fig:reduction}).
The {recursive} scheme is very convenient for explicit calculations
and contains the complete calculational chain of tensor reduction,
for both massive and massless propagators, and works with
dimensional regularization. The general case of tensor integrals
using dimensional regularization was treated in a series of papers
\cite{Bern:1992em,Bern:1993kr,Binoth:1999sp,Duplancic:2003tv,Diakonidis:2008ij},
thereby allowing also for massless particles.
The only necessary package to be added is one for the evaluation of
1-point to 4-point scalar integrals.
These are calculated by QcdLoop \cite{Ellis:2007qk}. This is a very
interesting property of the reduction, because using Fortran and
switching off the relative scalar master integrals we have all the
information that is needed to determine the coefficients of the scalar
master integrals that arise in calculations by unitarity methods.
In the same way, by switching off all the scalar
master integrals we can also extract the rational term. Finally, due to
the capability of QcdLoop to give the $1/\epsilon$ and $1/\epsilon^2$
terms, we can cross check the correctness of our loop amplitude as these
terms are proportional to the tree level amplitude.

The stripped amplitude in total is optimized by Maple.  We divide
the total number of diagrams to groups of hexagons, pentagons, boxes,
triangles, and bubbles, and then we translate the output to Maple. Maple
optimizes it and feeds it to a Fortran subroutine calculating the
amplitude contributions of the above groups. As mentioned above, these
structures are single colored ones. So each one of them is multiplied
with the appropriate color structure stored in the color subroutine.

Connecting all the pieces together, the stripped amplitudes including tensor
integrals and spinor lines separately calculated in specific subroutines,
the color structures precalculated and stored in an array, we calculate the
total loop amplitude. In the same way, using much simpler routines,
we calculate the tree level case.

The biggest advantage of the program is the fact that it is
flexible. With minor changes, we can implement it for
calculating different processes. We just have to provide Diana with the
process that we are interested in, fix the Feynman rules and be careful
with manipulation of possible spinor lines and colors. All the rest,
tensor reduction and optimization of the code are done automatically by
the program.

\section{Acknowledgments}
 Work supported in part by Sonderforschungsbereich/Trans\-re\-gio
SFB/TRR 9 of DFG
``Com\-pu\-ter\-ge\-st\"utz\-te Theoretische Teil\-chen\-phy\-sik"
and by the European Community's Marie-Curie Research Trai\-ning Network
MRTN-CT-2006-035505 ``HEPTOOLS''. We would also like to thank our
collaborators Jochem Fleischer and Tord Riemann for useful discussions.

\end{document}